\documentclass[%
 reprint,
superscriptaddress,
 amsmath,amssymb,
 aps,
]{revtex4-2}

\usepackage{graphicx}
\usepackage{dcolumn}
\usepackage{bm}
\usepackage{physics}
\usepackage{mathtools}
\usepackage{color}
\usepackage{changes}

\begin{document}

\preprint{APS/123-QED}

\title{Quantum nonlinear spectroscopy via correlations of weak Faraday-rotation measurements}

\author{Brian Chung Hang Cheung}
\affiliation{%
Department of Physics, The Chinese University of Hong Kong, Shatin, New Terrotories, Hong Kong, China}%
\author{Ren-Bao Liu}
\email{rbliu@cuhk.edu.hk}
\affiliation{%
Department of Physics, The Chinese University of Hong Kong, Shatin, New Terrotories, Hong Kong, China}%
\affiliation{Centre for Quantum Coherence, The Chinese University of Hong Kong, Shatin, New Terrotories, Hong Kong, China}%
\affiliation{The Hong Kong Institute of Quantum Information Science and Technology, The Chinese University of Hong Kong, Shatin, New Terrotories, Hong Kong, China}%
\affiliation{New Cornerstone Science Laboratory, The Chinese University of Hong Kong, Shatin, New Terrotories, Hong Kong, China}%

\date{\today}

\begin{abstract}
The correlations of fluctuations are key to studying fundamental quantum physics and quantum many-body dynamics. They are also useful information for understanding and combating decoherence in quantum technology. 
Nonlinear spectroscopy and noise spectroscopy are powerful tools to characterize fluctuations, but they can access only very few among the many types of higher-order correlations. 
A systematic quantum sensing approach, called quantum nonlinear spectroscopy (QNS), is recently proposed for extracting arbitrary types and orders of time-ordered correlations, using sequential weak measurement via a spin quantum sensor. However, the requirement of a central spin as the quantum sensor limits the versatility of the QNS since usually a central spin interacts only with a small number of particles in proximity and the measurement of single spins needs stringent conditions. 
Here we propose to employ the polarization (a pseudo-spin) of a coherent light beam as a quantum sensor for QNS. After interacting with a target system (such as a transparent magnetic material), the small Faraday rotation of the linearly polarized light can be measured, which constitutes a weak measurement of the magnetization in the target system.
Using a Mach-Zehnder interferometer with a designed phase shift, one can post-select the effects of the light-material interaction to be either a quantum evolution or a quantum measurement of the material magnetization. 
This way, the correlated difference photon counts of a certain numbers of measurement shots, each with a designated interference phase, can be made proportional to a certain type and order of correlations of the magnetic fluctuations in the material. 
The analysis of the signal-to-noise ratios shows that the second-order correlations are detectable in general under realistic conditions and higher-order correlations are significant when the correlation lengths of the fluctuations are comparable to the laser spot size (such as in systems near the critical points). 
Since the photon sensor can interact simultaneously with many particles and the interferometry is a standard technique, this protocol of QNS is advantageous for studying quantum many-body systems.

\end{abstract}

\maketitle


\section{\label{sec: I}Introduction}

The correlations of fluctuations are key to studying the foundation of quantum mechanics (such as via the Bell inequalities~\cite{PhysicsPhysiqueFizika.1.195,clauser1969proposed} and the Leggett-Garg inequalities~\cite{PhysRevLett.54.857}) and the dynamics and thermodynamics of quantum many-body systems~\cite{marconi2008fluctuation, sinitsyn2016noise_theory}. They are also useful information for characterizing the noises and environments in quantum technology~\cite{clausen2010bath,yang2011preserving,suter2016RMPQControl,Yang_2016}. The second-order correlations have been widely studied, such as in linear optics, transport, thermal response, magnetic response, and neutron scattering. Higher-order correlations are usually ignored by the assumption that they are weaker than second-order correlations, or they can be factorized into second-order correlations by Wick's theorem when the fluctuations are Gaussian (which is often the case in macroscopic systems  far away from the critical points)~\cite{Yang_2016}. 
However, there are important scenarios where the irreducible higher-order correlations are significant. One such case is the critical fluctuations near the phase transitions~\cite{joubaud2008experimental,bramwell2009distribution}. Recent studies also show the importance of high-order correlations in quantum sensing~\cite{Liu_2010, Wang_2021, Meinel_2022}, in mesoscopic quantum systems~\cite{schweigler2017experimental,Schweigler_2021,alvarez2015localization}, and in spin systems under external noises~\cite{li2016higher}. 
The characterization of higher-order correlations thus is desirable~\cite{Wang_2019, PhysRevA.98.042111, H_gele_2018}.

There is a rich structure in higher-order correlations of quantum fluctuations~\cite{Haehl_2019}.
Corresponding to different orderings of quantum operators, there are in general $K!$ independent correlations among $K$ different quantities $\hat{B}_1,\hat{B}_2,\ldots, \hat{B}_K$, given by, e.g.,
$\left\langle\hat{B}_{\sigma(1)}\hat{B}_{\sigma(2)}\cdots\hat{B}_{\sigma(K)}\right\rangle$ (with $\sigma$ denoting an element of the permutation group $S_K$ and $\langle\cdots\rangle$ denoting the average over the system density operator).
For $K$ quantities at different times (specifically, $\hat{B}_1,\hat{B}_2,\ldots, \hat{B}_K$ at $t_1,t_2,\ldots,t_K$ correspondingly with $t_1\le t_2\cdots\le t_K$), most of these correlations are the so-called out-of-time-order correlators (OTOCs), which do not directly affect the dynamics of the system. There are $2^{K-1}$ independent time-ordered correlations, corresponding to different arrangements of the operators $B_{k}$ on the two branches of the contour evolution (i.e.,  on the left or right side of the system density operator $\hat{\rho}$). The case for $K=2$ has been well studied in the form of Keldysh-Kadanoff-Baym Green's functions~\cite{keldysh1965diagram, kadanoff2018quantum}. The  time-ordered higher-order correlations can be formulated in an intriguing basis using the commutators or anti-commutators between the operators. 
We introduce the superoperators $\mathbb{B}^{\pm}_k$ at time $k$ with $\mathbb{B}^{+}_k$ corresponding to an anticommutator, defined as $\mathbb{B}^{+}_k\hat{\rho}\equiv \left(\hat{B}_k\hat{\rho}+\hat{\rho}\hat{B}_k\right)/2$, and $\mathbb{B}^{-}_k$ corresponding to a commutator, defined as $\mathbb{B}^{-}_k\hat{\rho}\equiv \left(\hat{B}_k\hat{\rho}-\hat{\rho}\hat{B}_k\right)/i$. 
The $2^{K-1}$ types of time-ordered $K$th-order correlations are $C^{\eta_K ... \eta_1}=\Tr{\mathbb{B}^{\eta_K}_K ... \mathbb{B}^{\eta_2}_2 \mathbb{B}^{\eta_1}_1 \hat{\rho}}$, where $\hat{\rho}$ is the initial state of the system and $\eta_k\in\{+,-\}$. Note that if $\eta_K=-$, the correlation always vanishes since the trace of a commutator is zero. For $\eta_K=+$, the correlations can also be written as
\begin{align}
    C^{+\eta_{K-1}\cdots\eta_1} & ={\rm Tr}\left[\hat{\rho} \left(\mathbb{B}^{\eta_1}_1\cdots \mathbb{B}^{\eta_{K-1}}_{K-1}\hat{B}_K\right)\right]\nonumber \\ & \equiv \eta_1\eta_2\cdots\eta_{K-1}\left\langle\mathbb{B}^{\eta_1}_1\cdots \mathbb{B}^{\eta_{K-1}}_{K-1}\hat{B}_K\right\rangle.
    \nonumber
\end{align}
This rich structure of correlations is important in the dynamic of quantum system~\cite{PhysRevA.98.042111} but however have been largely unexplored due to the limitation of conventional spectroscopy~\cite{mukamel1995principles,crooker2004spectroscopy,Liu_2010,laraoui2013corr,li2016higher,Crooker2014noiseBeyond,sinitsyn2016noise_theory}. 
Conventional nonlinear spectroscopy~\cite{mukamel1995principles}, using classical probes, measures only one of the $2^{K-1}$ type of correlations, namely, $C^{+-- ...-}=i^{K-1}\left\langle\left[\hat{B_1},\left[\hat{B}_2,\ldots\left[\hat{B}_{K-1},\hat{B}_K\right]\right]\right]\right\rangle$, corresponding to the $(K-1)$-th order nonlinear susceptibility, while noise spectroscopy usually measures only $C^{++...+}$~\cite{crooker2004spectroscopy,Liu_2010,laraoui2013corr,li2016higher,Crooker2014noiseBeyond, sinitsyn2016noise_theory}.

It is recently proposed that arbitrary types and orders of time-ordered correlations can be extracted by a systematic quantum sensing approach, coined quantum nonlinear spectroscopy (QNS), using the correlations of sequential weak measurement~\cite{Pfender_2019} via a spin quantum sensor~\cite{Wang_2019}. This scheme has been adopted in experiments to single out quantum signals from classical noises~\cite{Meinel_2022,shen2023PRL}. 
The quantum sensor interacts with the target system in a sequence of short periods, and the projective measurement of the sensor constitutes a weak measurement of the target. 
In the interaction picture, the coupled sensor-target system evolves by the quantum Liouville equation $\partial_t\hat{\rho}= -i\comm{\hat{V}}{\hat{\rho}}$ (hereafter the Planck constant $\hbar$ is set as unity), where $\hat{V}=\hat{S}\otimes\hat{B}$ with $\hat{S}$ being a sensor operator and $\hat{B}$ a target operator.  For a small period of interaction, starting from an initial state $\hat{\rho}=\hat{\rho}_{s}\otimes\hat{\rho}_B$ (with $\rho_{s/B}$ denoting the sensor/target state), the coupled system evolves to $\hat{\rho}(t+\tau)=\hat{\rho}-i\tau\comm{\hat{V}}{\hat{\rho}}+O\left(\tau^2\right)$. 
The commutator can be decomposed into
\begin{align}
i^{-1}\left[{\hat{S}\otimes\hat{B}},{\hat{\rho}_{s}\otimes\hat{\rho}_B}\right] =\mathbb{S}^{+}\hat{\rho}_{s}\otimes\mathbb{B}^{-}\hat{\rho}_{B} +\mathbb{S}^{-}\hat{\rho}_{s}\otimes\mathbb{B}^{+}\hat{\rho}_{B}.
\label{eq:decomposition}
\end{align}
The physical meanings of the two terms in the r.h.s. of Eq. (\ref{eq:decomposition}) are clear:
\begin{enumerate}
    \item Considering $\Tr \mathbb{S}^{+}\hat{\rho}_{s}=\langle \hat{S}\rangle$, the first term corresponds to the target evolving (indicated by the commutator $\mathbb{B}^{-}_j\hat{\rho}_{B}$) under a quantum force  $\hat{S}$ from the sensor.
    \item Similarly, the second term is the evolution of the sensor under the quantum force $\hat{B}$ from the target, that is, the back-action.  
\end{enumerate}
After the interaction, the measurement of a physical quantity $\Lambda_j$ of the quantum sensor at time $t_j$ leads to a reduced density matrix of the target in the form of 
\begin{align}
    {\mathbb M}_j\hat{\rho}_B(t_j)
    \equiv & \Tr_{\rm s}\left[{\hat{\Lambda}_j\hat{\rho}}(t_j+\tau)\right] \nonumber \\ = & \Tr_{\rm s}\left(\hat{\Lambda}_j\hat{\rho}_{\rm s}\right) \hat{\rho}_{B}+\tau\Tr_{\rm s}\left(\hat{\Lambda}_j\mathbb{S}^{-}\hat{\rho}_{s}\right) \mathbb{B}^{+}_j\hat{\rho}_{B}\nonumber \\ & +\tau\Tr_{\rm s}\left(\hat{\Lambda}_j \mathbb{S}^{+}\hat{\rho}_{s}\right)\mathbb{B}^{-}_j\hat{\rho}_{B}+O\left(\tau^2\right).
\end{align}
The correlation of the physical quantities $\{\Lambda_1, \Lambda_2, ...,\Lambda_K\}$ measured at time $t_1, t_2, \ldots, t_K$ can be written as
\begin{equation} \label{eq1}
G^{(K)}=\Tr_B{(\mathbb{M}_{K}...\mathbb{M}_{2}\mathbb{M}_{1}\hat{\rho}_{B}(t_0)}).
\end{equation}
The initial state and measurement basis of the sensor can be chosen for the pre- and post-selection of the effect of the sensor-target interaction in each measurement period. For example, by choosing the initial state of the sensor $\hat{\rho}_{s,j}$ and the sensor quantity to be measured $\Lambda_j$, one can make ${\rm Tr}_{\rm s}\left[\hat{\Lambda}_j\hat{\rho}_{s,j}\right]=0$ and ${\rm Tr}_{\rm s}\left[\hat{\Lambda}_j\mathbb{S}^-\hat{\rho}_{s,j}\right]=0$, so that the weak measurement yields an evolution of the target, that is, $\mathbb{B}^-\hat{\rho}_{B}(t_j)$.
This way, the QNS realized by sequential weak measurement via a quantum sensor can extract arbitrary types and orders of the time-ordered correlations in a quantum system.

However, the requirement of a central spin as the quantum sensor limits the versatility of the QNS since usually a central spin interacts only with a small number of particles in proximity and the measurement of single spins demands stringent conditions.
The polarization (a pseudo-spin) of a coherent light beam can be an alternative for a central spin as a quantum sensor~\cite{Liu_2010, Chen_2014}. The light beam as a quantum sensor has the advantage of large coherent length (so that it can interact with a quantum many-body system) and has the ease for preparation, control and detection. The pseudo-spin of light has been routinely used in noise spectroscopy of spin systems~\cite{crooker2004spectroscopy,crooker2004spectroscopy,song2022collision}.

In this paper, we present a scheme of QNS using polarization of a coherent light beam as a quantum sensor.
The interaction with the target system (such as spins in a transparent material) results in small Faraday rotation of the linearly polarized light. The measurement of the Faraday rotation constitutes a weak measurement of the spins. The Faraday rotation can be measured with a Mach-Zehnder interferometer with a phaseshift chosen such that the effects of the light-material interaction is selected to be either a quantum evolution or a quantum measurement of the magnetization of the spins (corresponding to a commutator or anti-commutator between the magnetizaiton and the density operator of the spins). As a result, the correlated photon counts of a sequence of measurement shots, each with a designated phaseshift, can be made proportional to a designated type and order of correlation of the spin fluctuations. Analysis of the signal-to-noise ratio for the measurement of  correlations shows that the scheme is feasible under realistic conditions.

\section{\label{sec:II}Scheme}

\subsection{Basic idea}
\begin{figure}[h]
    \centering
    \includegraphics[width=0.9\columnwidth ]{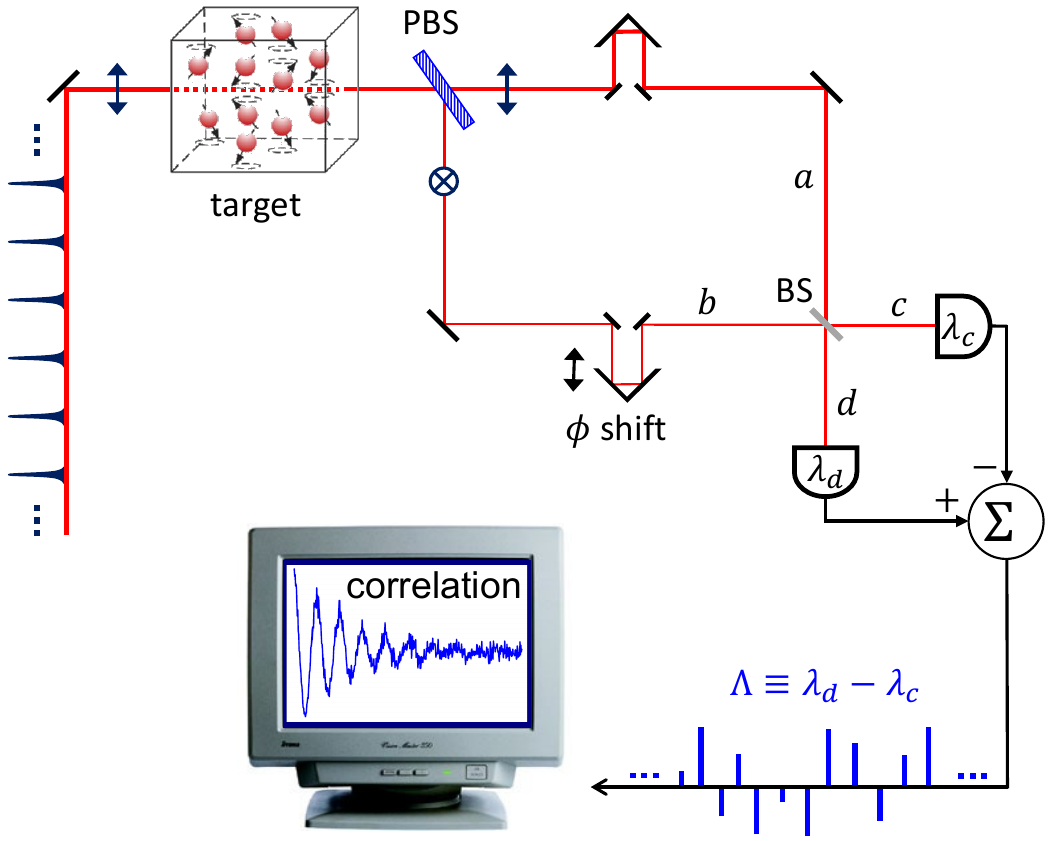}
    \caption{Schematic of a Mach-Zehnder interferometer for detecting arbitrary correlations of a quantum spin system using a polarized light beam sensor. The difference photon count $\Lambda$ is detected to measure the polarization of the light after the interaction with the target system. A designated phase shift $\phi$ is induced by a variable optical delay to select the basis of polarization measurement.}
     \label{fig:1}
\end{figure}

We employ the Mach-Zehnder interferometer in which the polarization of a light beam can be measured in a designated basis by choosing a certain relative phaseshift between the two arms (see Fig.~\ref{fig:1}). A sequence of linearly polarized laser pulses, described by the coherent state $\dyad{\alpha, H}$  (with $H/V$ denoting horizontal/vertical polarization), are applied to a target system. The light polarization acts as the quantum sensor, coupled to a spin system (the target). The interaction with the spins leads to a Faraday rotation of the light polarization. After the interaction, the light beam is separated into two arms by a polarizing beam splitter (PBS), such that the component with horizontal polarization is reflected and the component with vertical polarization is transmitted. 
A phaseshift $\phi$ is then applied in one arm (e.g., the reflected beam) and then the two arms are mixed through a 1:1 beam splitter (BS). The difference between the photon counts in the two output directions is recorded to measure the light polarization, in a basis that is selected by choosing the phase shift $\phi$. As the light polarization is entangled with the target spins, the measurement of the light polarization constitutes an indirect measurement of the target spins. When the interaction between the light polarization and the spins is weak, i.e., the Faraday rotation is small, the indirect measurement on the spins is a weak one.   
The correlation among a sequence of difference photon counts corresponds to a certain correlation in the target spin system.

\subsection{\label{sec:II A}Light polarization as a pseudo-spin-1/2 sensor}
We can define a pseudo-spin-1/2 using Jones vectors. In the formalism of Jones vectors, we write the horizontal (H) and vertical (V) linear polarizations as
$$|H\rangle\equiv\left(\begin{array}{c} 1 \\ 0\end{array}\right), \ \ |V\rangle\equiv\left(\begin{array}{c} 0 \\ 1 \end{array}\right),$$
the diagonal (D) and anti-diagonal (A) linear polarizations as
$$|D\rangle\equiv\frac{1}{\sqrt{2}}\left(\begin{array}{c} 1 \\ 1\end{array}\right),\ |A\rangle\equiv\frac{1}{\sqrt{2}}\left(\begin{array}{c} +1 \\ -1\end{array}\right), $$
and the two circular polarizations as
$$  |R\rangle\equiv \frac{1}{\sqrt{2}}\left(\begin{array}{c} +1 \\ -i\end{array}\right), \ |L\rangle\equiv\frac{1}{\sqrt{2}}\left(\begin{array}{c} +1 \\ +i\end{array}\right).$$
In analogue to a spin-1/2, we define three photon operators
\begin{subequations}
\begin{align}
\hat{S}_1 &\equiv \frac{\hat{n}_H-\hat{n}_V}{2}= \frac{1}{2}\left(\hat{a}_H^{\dag},\hat{a}_V^{\dag}\right)
\left(\begin{array}{cc} +1 & 0 \\ 0 & -1\end{array}\right) \left(\begin{array}{c}\hat{a}_H \\ \hat{a}_V\end{array}\right), \nonumber \\
\hat{S}_2 &\equiv \frac{\hat{n}_D-\hat{n}_A}{2}= \frac{1}{2} \left(\hat{a}_H^{\dag},\hat{a}_V^{\dag}\right)
\left(\begin{array}{cc} 0 & +1  \\  +1 & 0 \end{array}\right) \left(\begin{array}{c}\hat{a}_H \\ \hat{a}_V\end{array}\right), \nonumber \\
\hat{S}_3 &\equiv \frac{\hat{n}_R-\hat{n}_L}{2}= \frac{1}{2} \left(\hat{a}_H^{\dag},\hat{a}_V^{\dag}\right)
\left(\begin{array}{cc} 0 & -i \\ +i & 0 \end{array}\right) \left(\begin{array}{c}\hat{a}_H \\ \hat{a}_V\end{array}\right), \nonumber
\end{align}
\end{subequations}
where $\hat{a}_{\alpha}$ is the photon annihilation operator for the polarization $\alpha$ and $\hat{n}_{\alpha}\equiv \hat{a}_{\alpha}^{\dag}\hat{a}_{\alpha}$ is the photon number operator. The pseudo-spin operators satisfy the same commutators as the spin operators, namely,
\begin{align}
\left[\hat{S}_{a},\hat{S}_{b}\right]=i\epsilon_{abc}\hat{S}_{c}.
\end{align}
But they do not have the same anti-commutators as the spin-1/2. For example,
\begin{subequations}
\begin{align}
\left\{\hat{S}_2,\hat{S}_3\right\} & =\frac{i}{2}\hat{a}_V^{\dag}\hat{a}_V^{\dag}\hat{a}_H\hat{a}_H+{\rm H.c.} \ne 0 , \\
\left\{\hat{S}_3,\hat{S}_3\right\} & =\hat{n}_H\hat{n}_V+\frac{\hat{n}_H+\hat{n}_V}{2}-\frac{\left(\hat{a}^{\dag}_H\hat{a}_V\right)^2+{\rm H.c.}}{2}\ne \frac{1}{2}.
\end{align}
\end{subequations} 

\subsection{\label{sec:II B} Single Shot of measurement}

Let us first consider a single shot of measurement. In the interaction picture, the evolution of the coupled system of photons and spins is determined by 
$$\partial_t\hat{\rho}(t)=-i\comm{\hat{V}(t)}{\hat{\rho}},$$
with the spin-photon interaction
\begin{align}\hat{V}(t)=\hat{S}_3\otimes\hat{B}(t)=-\frac{i}{2}(\hat{a}^{\dagger}_{H}\hat{a}_{V}-\hat{a}^{\dagger}_{V}\hat{a}_{H})\otimes\hat{B}(t),\end{align}
where the target operator $\hat{B}$ is an effective field proportional to the magnetization of the spins along the light propagation direction.
The magnetization induces opposite frequency shifts for the left and right circularly polarized photons and hence opposite phase shifts, which causes the Faraday rotation of the linear polarization.

We assume the pulse duration $\tau$ is small, that is, the Faraday rotation angle $||\hat{B}\tau||\ll 1 $ and $\hat{B}(t)\approx\hat{B}(t+\tau)$. After the interaction for a short period of time $\tau$, starting from the initial state $\hat{\rho}(t)=\hat{\rho}_{\rm s} \otimes\hat{\rho}_B$, the coupled system evolves to the state
\begin{align}
\hat{\rho}(t+\tau)\approx  \hat{\rho}(t) & +\tau \mathbb{S}_3^{-}\hat{\rho}_{s}\otimes\mathbb{B}^{+}\hat{\rho}_{B}
 +\tau \mathbb{S}_3^{+}\hat{\rho}_{s}\otimes\mathbb{B}^{-}\hat{\rho}_{B}.
\label{eq3}
\end{align}

The physical meaning of the term $\tau \mathbb{S}_3^{-}\hat{\rho}_{s}\otimes\mathbb{B}^{+}\hat{\rho}_{B}$ in Eq.~(\ref{eq3}) is the pseudo-spin rotating about the 3rd axis under the field $\hat{B}$ from the target. To select this effect, we can prepare the pseudo-spin initially in an eigenstate of $\hat{S}_1$ and then measure the photon polarization in the basis of $\hat{S}_2$, that it, in the basis of the linear polarizations $D$ and $A$. Indeed, if the sensor initial state is chosen as
$$\hat{\rho}_{\rm s}(t_j)=|\alpha,H\rangle\langle\alpha,H|,$$
(with $|\alpha,H\rangle\equiv \exp\left(\alpha\hat{a}_H^{\dag}-\alpha^*\hat{a}_H\right)|{\rm vacuum}\rangle$) and the observable to be measured as $\hat{\Lambda}_j=\hat{S}_2$, we have
$\Tr_{\rm s} \left[ \hat{S}_2\hat{\rho}_{\rm s}(t_j)\right]=0$, $\Tr_{\rm s}\left[\hat{S}_2\mathbb{S}_3^{+}\hat{\rho}_{s}(t_j)\right]=\frac{1}{2}\left\langle\left\{\hat{S}_2,\hat{S}_3\right\}\right\rangle=0$, and 
$\Tr_{\rm s}\left[\hat{S}_2\mathbb{S}_3^{-}\hat{\rho}_{s}(t_j)\right]=\frac{1}{i}\left\langle\left[\hat{S}_2,\hat{S}_3\right]\right\rangle=\frac{1}{2}\left|\alpha\right|^2$. Therefore, the measurement induces the superoperator on the bath as
\begin{align}
{\mathbb M}_j\hat{\rho}_B(t_j)=\frac{1}{2}\tau \left|\alpha\right|^2 {\mathbb B}_j^{+}\hat{\rho}_B(t_j), 
\ {\rm for}\ \hat{\Lambda}_j=\hat{S}_2.
\label{eq:S1S2}
\end{align}
The physical meaning of the other term $\tau \mathbb{S}_3^{+}\hat{\rho}_{s}\otimes\mathbb{B}^{-}\hat{\rho}_{B}$ in Eq.~(\ref{eq3}) is the quantum force $\hat{S}_3$ from the sensor driving the evolution of the target. To select this effect, we should measure the pseudo-spin component $\hat{S}_3$. Using the same initial state $\hat{\rho}_{\rm s}(t_j)=|\alpha,H\rangle\langle\alpha,H|$ and similar algebra for the derivation of Eq.~(\ref{eq:S1S2}), we get 
\begin{align}
{\mathbb M}_j\hat{\rho}_B(t_j)=\frac{1}{2}\tau \left|\alpha\right|^2 {\mathbb B}_j^{-}\hat{\rho}_B(t_j), 
\ {\rm for}\ \hat{\Lambda}_j=\hat{S}_3.
\label{eq:S1S3}
\end{align}

Using the scheme illustrated in Fig.~\ref{fig:1}, we can measure either $\hat{S}_2$ or $\hat{S}_3$ by designing the phase shift in the reflected path.
Interference occurs between transmitted path and reflected path. The 1:1 BS transforms the input photon operators $\hat{a}$ and $\hat{b}$ into output ones $\hat{c}$ and $\hat{d}$ by
\begin{equation*}
    \begin{pmatrix}
    \hat{c} \\
    \hat{d} 
    \end{pmatrix}=
    \frac{1}{\sqrt{2}}
    \begin{pmatrix}
        1 & i\\
        i & 1
    \end{pmatrix}
    \begin{pmatrix}
    \hat{a} \\
    \hat{b} 
    \end{pmatrix}.
\end{equation*}

For the measurement of $\hat{S}_3$, we set the phase shift such that $e^{i\phi}=1$ and therefore the input annihilation operators are $\hat{a}=\hat{a}_H$ and $\hat{b}=\hat{a}_V$. Then, the outputs become
\begin{equation*}
     \frac{1}{\sqrt{2}}
    \begin{pmatrix}
        1 & i\\
        i & 1
    \end{pmatrix}
    \begin{pmatrix}
    \hat{a}_H \\
    \hat{a}_V 
    \end{pmatrix}=  \begin{pmatrix}
    \hat{a}_L \\
    i\hat{a}_R 
    \end{pmatrix},
\end{equation*}
and the difference photon count $\Lambda\equiv\lambda_d-\lambda_c$ as shown in Fig.~\ref{fig:1} becomes $\hat{a}_R^{\dagger}\hat{a}_R-\hat{a}_L^{\dagger}\hat{a}_L\equiv\hat{S}_3$.

Similarly, for the measurement of $\hat{S}_2$, we set the phase shift such that $e^{i\phi}=i$ and the input annihilation operators are $\hat{a}=\hat{a}_H$ and $\hat{b}=i\hat{a}_V$.
Then, the outputs become
\begin{equation*}
      \frac{1}{\sqrt{2}}
    \begin{pmatrix}
        1 & i\\
        i & 1
    \end{pmatrix}
    \begin{pmatrix}
    \hat{a}_H \\
    i\hat{a}_V 
    \end{pmatrix}= \begin{pmatrix}
    \hat{a}_A \\
    i\hat{a}_D 
    \end{pmatrix},
\end{equation*}
and the difference photon count $\Lambda\equiv \lambda_d-\lambda_c$ becomes $\hat{a}_D^{\dagger}\hat{a}_D-\hat{a}_A^{\dagger}\hat{a}_A\equiv\hat{S}_2$.

\subsection{\label{sec: II C} Correlations of sequential weak measurements}

The correlation of a sequence of $K$ shots of measurement is
\begin{align} 
G^{(K)}=\Tr_B\left[{\mathbb M}_K\cdots{\mathbb M}_2{\mathbb M}_1\hat{\rho}_B\right].
\label{eq:GK}
\end{align}

Using Eqs.~(\ref{eq:S1S2}) and (\ref{eq:S1S3}), we find that the correlated photon count differences $G^{(K)}$ is proportional to a certain type of correlation in the target system as
\begin{equation}
    G^{(K)}=2^{-K} \tau^K \abs{\alpha}^{2K}C^{\eta_K...\eta_2\eta_1}
\label{eq: GK_C_relation}
\end{equation}
where $\eta_j=+$ or $-$ if the $j^{th}$ shot of measurement is chosen as $\hat{\Lambda}_j=\hat{S}_{2}$ or $\hat{S}_3$, respectively. 
For example, choosing $\hat{\Lambda}_2=\hat{S}_2$ and $\hat{\Lambda}_1=\hat{S}_3$, we obtain 
\begin{equation*}
    G^{(2)}=2^{-2} \tau^2 \abs{\alpha}^{4}C^{+-},
\end{equation*}
and choosing $\hat{\Lambda}_4=\hat{S}_2$, $\hat{\Lambda}_3=\hat{S}_3$, $\hat{\Lambda}_2=\hat{S}_3$ and $\hat{\Lambda}_1=\hat{S}_2$, we get 
\begin{equation*}
    G^{(4)}= 2^{-4} \tau^4 \abs{\alpha}^{8}C^{+--+}.
\end{equation*}
This way we can obtain arbitrary types and orders of correlations just by getting the correlated photon count differences from the interferometer.

\section{\label{sec:level1} Signal-to-noise ratio}
We calculate the signal-to-noise (SNR) ratio. We consider the shot noises~\cite{Giovannetti_2011} of the photon counts and neglect the technical noises such as the laser fluctuations and the imperfections in beam splitting, phase shift, and photon detection, since these factors would highly depend on the specific systems employed in experiments. The SNR is calculated for $L$ repeated sequences of $K$ shots of measurement for obtaining a $K$-th order correlation $G^{(K)}$.

\subsection{\label{sec: V A}SNR for first-order correlations}

The signal $\expval{S}$ for a first-order correlation $G^{(1)}$ is the average difference photon count between the two outputs, 
\begin{equation*}
    \expval{S}=G^{(1)}=\frac{1}{2}\tau \abs{\alpha}^{2}C^{+}.
\end{equation*}
The variance of signal is $\sigma^2=\expval{S^2}-\expval{S}^2$. 
Since $\expval{S}^2$ is of $\order{\tau^2}$, which is much smaller than $\expval{S^2}$, the variance can be simplified as $$\sigma^2\approx\expval{S^2}=\expval{\Lambda^2}=\Tr[\hat{S}_{2}^2\hat{\rho}(t+\tau)].$$ 
Using Eq.~(\ref{eq3}) and keeping the leading-order terms, we get 
\begin{equation}   \sigma^2\approx\Tr\left[\hat{S}_{2/3}^2\hat{\rho}(t)\right]=\abs{\alpha}^{2}.
\end{equation}
Therefore, the SNR for first-order correlations in the shot-noise limit is 
\begin{equation}
    {\rm SNR}^{(1)}=\frac{L\expval{S}}{\sqrt{L\sigma^2}}=\frac{\sqrt{L}}{2}\abs{\alpha}\tau C^{+}
\end{equation}

\subsection{\label{sec: V B}SNR for $K$-th order correlations}

Similarly, the signal of measuring a $K$-th order correlation given by Eq.~(\ref{eq: GK_C_relation}) is $\expval{S}=G^{(K)}$. 
In the leading order, the variance $\sigma^2\approx \expval{S^2}=\expval{\Lambda_K^2\cdots\Lambda_2^2\Lambda_1^2}=\expval{\Lambda_K^2}\cdots\expval{\Lambda_2^2}\expval{\Lambda_1^2}$, where the subscript in $\Lambda_k$ labels the shot of measurement.  
By straightforward calculation, we get $\expval{\Lambda_k^2} \approx \abs{\alpha}^2$ in the leading order. Thus, the variance of measuring a $K$-th order correlation is
$$     \expval{S^2}\approx \abs{\alpha}^{2K}.$$
Therefore, the SNR for a $K$th order correlation is 
$$
    {\rm SNR}^{(K)}=\sqrt{L}\frac{\expval{S}}{\sigma}=2^{-K}\sqrt{L}\abs{\alpha}^{K} \tau^{K}C^{\eta_K...\eta_2\eta_1}.
$$

\subsection{\label{sec:VI} Estimation of the signal strength}

We consider a target system containing an ensemble of spins.  We assume that the laser pulse length is much greater than the thickness of the sample ($D$) and the laser pulse duration $\tau$ is much shorter than the timescale of the spin dynamics. In this case, all the spins $\{ \hat{\mathbf J}_i |i=1,2,\ldots, N_{\rm s} \}$ within the laser spot interact simultaneously with the photons. The Faraday rotation $\hat{\theta}(t)$ is related to the magnetization $\hat{J}(t)\equiv N_{\rm s}^{-1}\sum_{i}\hat{J}^z_i$ along the light propagation direction (the $z$-axis) by~\cite{Chen_2014}
\begin{equation}
\hat{\theta} (t)=g  D \hat{J}(t)=\tau\hat{B}(t),
\end{equation}
where $g$ is the coupling constant. 
The $K$-th order correlation is related to the correlation of the spins by
\begin{equation}
\tau^KC^{\eta_K\cdots\eta_2\eta_1}=g^KD^K\Tr({\mathbb J}^{\eta_k}_K\cdots{\mathbb J}_2^{\eta_2}{\mathbb J}_1^{\eta_1}\hat{\rho}_B) .
\end{equation}
When the spins are uncorrelated, each spin contributes to the correlation independently, and therefore the $K$th order correlation of the $N_{\rm s}$ spins is estimated to be
$$
\tau^KC^{\eta_K\cdots\eta_2\eta_1}\sim \left(\frac{g D}{N_{\rm s}}\right)^K N_{\rm s}\expval{\left(\hat{J}_i^z\right)^K}.
$$
The SNR is
\begin{equation}
{\rm SNR}^{(K)}\sim \sqrt{L}\left(\frac{g N_{\rm ph}^{1/2}}{2 n_{\rm s} A}\right)^K n_{\rm s} D A\expval{\left(\hat{J}_i^z\right)^K},
\end{equation}
where $N_{\rm ph}=\abs{\alpha}^2$ is the average number of photons per laser pulse, $n_{\rm s}$ is the spin density of the target system, and $A$ is the area of the laser spot on the sample. To ensure the higher-order correlations be observable (for $K>2$), the laser should be strong enough and the light beam should be well focused such that  ${g N_{\rm ph}^{1/2}}/\left({2 n_{\rm s} A}\right)\gtrsim 1$.

To evaluate the feasibility of the measurement, we consider LiHoF$_4$~\cite{battison1975ferromagnetism} as a specific material system. This material is
a transparent Ising spin system with $\left(\hat{J}_i^z\right)^2= 8^2$ and presents interesting quantum phase transitions of magnetization~\cite{ronnow2005quantum}. Using the data in Ref.~\cite{battison1975ferromagnetism}, the coupling constant for a laser with wavelength about 435~nm (near resonant with the band gap of the material) is $g\approx 20$~rad/cm. The density of spins of LiHoF$_4$ is $n_{\rm s}\approx 1.39\times 10^{28}$~cm$^{-3}$.  Considering a laser pulse with focus size $\sim 10^{-4} $~cm and the number of photons $n_{\rm ph}\sim 10^{14}$ and a sample of $1$~cm thickness, we estimate the SNR as
$$
{\rm SNR}^{(K)}\sim \sqrt{L}\left(8\times 10^{-12}\right)^K \times 10^{20}.
$$
To get SNR$>1$, the number of measurement shots should be $L\gtrsim 10^{5}$ for measuring the second order correlations, which is quite feasible. For measuring the fourth order correlations ($K=4$), however, the number of measurement shots should be $L>10^{49}$, which is impossible for any practical experiments. This difficulty in measuring the higher-order correlations arises from the fact that uncorrelated or non-interacting spin ensembles typically exhibit negligible higher-order correlations.

An interesting scenario is that the higher-order correlations in interacting spin ensembles become significant near the critical points where the spins fluctuate collectively with a diverging correlation length $\xi$. The spins within a region of size $\xi$ can be taken as a large spin ($\sim n_{\rm s}\xi^3 J$). Thus, considering the collective fluctuations, the SNR for measuring the correlations is modified to be
\begin{equation}
{\rm SNR}^{(K)}\sim \sqrt{L}\left(\frac{g \xi^3 N_{\rm ph}^{1/2}}{2 A}\right)^K \frac{D A}{\xi^3}\expval{\left(\hat{J}_i^z\right)^K}.
\end{equation}
When the system is close to the critical point such that the correlation length is comparable to the size of the laser focus spot, the measurement of higher order correlations becomes feasible. Thus, the scheme presented in this paper provides an approach to studying the effects of higher-order correlations in interacting spin systems near the critical point.

\section{\label{sec:VII} Conclusion}
In summary, we propose an approach to selectively extracting arbitrary types and
orders of quantum correlations via weak Faraday rotation measurement. Unlike the single spin sensors used in existing  schemes~\cite{Wang_2019,Wang_2021,shen2023PRL,Meinel_2022,wu2022detection}, which interact with only a few spins in proximity, the quantum sensor employed in this scheme, namely, the optical polarization as a pseudo-spin, can couple uniformly to a quantum many-body system.
We have analysed the signal-to-noise ratio to show the feasibility of the scheme. While the detection of second-order correlations of fluctuations suffices to characterize quantum many-body systems far away from critical points, the higher-order correlations could be important near the critical points. When the systems approach the critical points, the collective fluctuations have diverging correlation lengths. The optical detection of higher-order correlations is feasible when the correlation lengths are comparable to the focus size of the laser beam.
The scheme proposed in this paper belongs to the category of quantum nonlinear spectroscopy~\cite{Meinel_2022}, for its capability of extracting the rich structure of higher-order quantum correlations, which are inaccessible to conventional nonlinear optical spectroscopy~\cite{mukamel1995principles}. The quantum nonlinear spectroscopy is not only a new tool for studying quantum many-body physics, but also new approaches to quantum science and technology, such as the study of quantum foundation (higher-order Leggett-Garg inequalities~\cite{PhysRevLett.54.857} and Bell inequalities~\cite{PhysicsPhysiqueFizika.1.195,clauser1969proposed}), optimal quantum control~\cite{clausen2010bath,yang2011preserving,Liu_2013,Viola_1999}, and quantum sensing~\cite{Wang_2021, Meinel_2022,shen2023PRL}).

\begin{acknowledgments}

This work was supported by the National Natural Science Foundation of China/Hong Kong Research Council Collaborative Research Scheme Project CRS-CUHK401/22, the Hong Kong Research Grants Council  Senior Research Fellow Scheme Project SRFS2223-4S01, and the New Cornerstone Science Foundation.

\end{acknowledgments}

\nocite{*}

\bibliography{Faraday_QNS}

\end{document}